\documentclass[
reprint,
superscriptaddress,
amsmath,amssymb,
aps, prl
]{revtex4-1}

\usepackage[dvipdfmx]{graphicx} 
\usepackage{dcolumn}
\usepackage{bm}

\begin{document}

\preprint{APS/123-QED}

\title{Chaotic synchronization induced by external noise in coupled limit cycle oscillators}

\author{Keiji Okumura}
\email{kokumura@ier.hit-u.ac.jp}
\affiliation{
 Institute of Economic Research,
 Hitotsubashi University,
 2--1 Naka, Kunitachi, Tokyo 186--8603, Japan
 }
\author{Akihisa Ichiki}
\affiliation{
 Institutes of Innovation for Future Society,
 Nagoya University,
 Furo--cho, Chikusaku, Nagoya 464--8603, Japan
}


\date{\today}

\begin{abstract}
A solvable model of noise effects on globally coupled limit cycle oscillators is proposed. The oscillators are under the influence of independent and additive white Gaussian noise. The averaged motion equation of the system with infinitely coupled oscillators is derived without any approximation through an analysis based on the nonlinear Fokker--Planck equation. Chaotic synchronization associated with the appearance of macroscopic chaotic behavior is shown by investigating the changes in averaged motion with increasing noise intensity. 
\end{abstract}

\maketitle


The discovery of the synchronization phenomena of chaos was a remarkable one since it seems at first glance to conflict with the definition of chaos. The stability of complete synchronization in coupled chaotic systems was studied \cite{Fujisaka_Stability_1983}. It was also shown that chaotic synchronization emerges in unidirectionally coupled oscillators in a model and an electronic circuit \cite{Pecora_Synchronization_1990}. In such a drive--response system, a functional relation was captured as generalized synchronization in refinement \cite{Rulkov_Generalized_1995}. The comprehensive understanding of synchronization transitions among complete, phase, and lag synchronizations was achieved using chaotic systems with a slight parameter mismatch \cite{Rosenblum_Phase_1996,Rosenblum_From_1997}. Recently, relationships among various types of chaotic synchronization were studied in a system of indirectly coupled oscillators with a common environment \cite{Resmi_Synchronized_2010}.

Sometimes, noise plays a counterintuitive role; in the case of uncoupled chaotic oscillators, it is widely known that common noise inputs can induce chaotic synchronization \cite{Toral_Analytical_2001}. However, much discussion was held at the time of the discovery of noise-induced chaotic synchronization (NICS) \cite{Maritan_Chaos_1994}. Some of these discussions were based on the following views: (i) NICS was a numerical artifact since the results were dependent on calculation accuracy \cite{Pikovsky_Comment_1994}, and (ii) the nonzero mean property of noise, which affected the system as a constant, would cause NICS \cite{Herzel_Chaos_1995}. Cumulative evidence associated with NICS clarified these issues \cite{Boccaletti_The_2002}. Once NICS was confirmed to be a reality, the relation between critical noise strengths and the size of the attractors \cite{He_Noise_2003}, NICS in phase \cite{Zhou_Noise_2002a}, effects of noise on generalized synchronization \cite{Guan_Effect_2006,Moskalenko_Effect_2011}, and NICS by colored noise were reported \cite{Yoshimura_Synchronization_2007}.

\begin{figure}[tb]
  \centering
    \includegraphics[width=5.5cm]{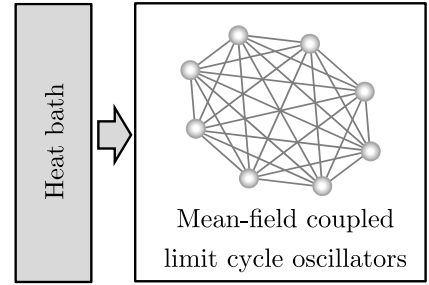}
    \caption{Schematic illustration of our system. 
    }
    \label{fig:Fig_schematic}
\end{figure}

The effects of noise on synchronization have also been studied in coupled oscillatory systems.　While noise can ``enhance'' the degree of chaotic synchronization constructively \cite{Zhou_Noise_2002b,Kiss_Noise_2003}, few studies reported chaotic synchronization purely ``induced'' by external noise in coupled oscillatory systems. Therefore, it is quite natural to ask whether NICS exists in coupled oscillators or not. To answer this fundamental academic question, in this letter, a solvable model for independent external noise is proposed to reveal the occurrence of NICS in mean-field coupled oscillatory systems, as shown in Fig.~\ref{fig:Fig_schematic}. It is worth noting that the study of NICS observed in coupled oscillatory systems solvable for external noise would contribute to resolving the complicated aspects mentioned before.

A potential approach to identify the occurrence of NICS in coupled oscillators would be related to the emergence of macroscopic chaotic behavior as an average. Let us suppose that collective motions described by an order parameter system show limit cycle oscillation under the presence of noise. In such cases, the oscillators exhibit roughly synchronous behavior to realize the limit cycle oscillation as an average though they are disturbed by noise. It would correspond to complete synchronization under the influence of noise \cite{Pikovsky_Synchronization_2001}. In the same way, if an order parameter behaves chaotically, it is suggestive of chaotic synchronization under the existence of noise. Thus, the occurrence of chaotic behavior in the order parameter system owing to the external noise, which is nonchaotic in the deterministic limit, would correspond to NICS.

\begin{figure}[tb]
  \centering
    \includegraphics[width=8.6cm]{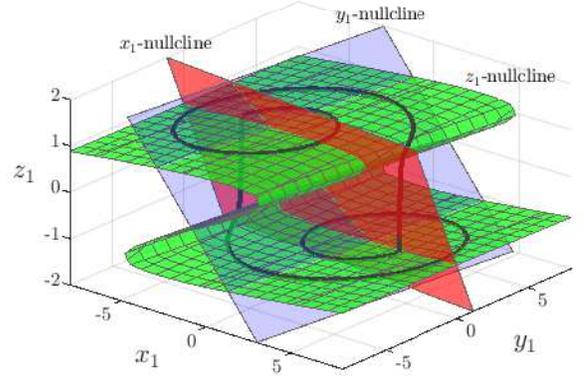}
    \caption{(Color online) Dynamical behavior of the single limit cycle oscillator without noise. 
    The trajectory for sufficiently large times is depicted. 
    The red and blue planes and the green curve correspond to the $x_{1}$-, $y_{1}$-, and $z_{1}$-nullclines, respectively. 
    The parameter values are 
    $a^{(x)}=1.0$, $b^{(x)}=1.1$, $c^{(x)}=d^{(x)}=1.0$, $J^{(x)}=1.0$, 
    $a^{(y)}=1.0$, $b^{(y)}=-1.0$, $c^{(y)}=1.1$, $d^{(y)}=-1.9$, $J^{(y)}=1.0$, 
    $a^{(z)}=100.0$, $b^{(z)}=-0.1$, $c^{(z)}=0.01$, $d^{(z)}=1.0$, and $J^{(z)}=200.0$.
    }
    \label{fig:Fig_3D_dG_DS_BDL}
\end{figure}

To reveal the existence of NICS in globally coupled limit cycle oscillators, let us introduce the following form of the Langevin equations, which is solvable for external noise \cite{Shiino_Chaos_2001,Ichiki_Chaos_2007,Okumura_Stochastic_2010}. For site $i$ $(i =1,...,N)$, it is given as 
\begin{eqnarray}
\dot{x}_{i} &=& 
- a^{(x)}x_{i} + \displaystyle \frac{J^{(x)}}{N}\sum_{j=1}^{N}F^{(x)}(b^{(x)}x_{j} + c^{(x)}y_{j} + d^{(x)}z_{j}) \nonumber\\
&&\mbox{} + \eta_{i}^{(x)}(t), 
\label{eq:3D_x}
\end{eqnarray}
\begin{eqnarray}
\dot{y}_{i} &=& 
- a^{(y)}y_{i} + \displaystyle \frac{J^{(y)}}{N}\sum_{j=1}^{N}F^{(y)}(b^{(y)}x_{j} + c^{(y)}y_{j} + d^{(y)}z_{j}) \nonumber\\
&&\mbox{} + \eta_{i}^{(y)}(t), 
\label{eq:3D_y}\\
\dot{z}_{i} &=& 
- a^{(z)}z_{i} + \displaystyle \frac{J^{(z)}}{N}\sum_{j=1}^{N}F^{(z)}(b^{(z)}x_{j} + c^{(z)}y_{j} + d^{(z)}z_{j}) \nonumber\\
&&\mbox{} + \eta_{i}^{(z)}(t), 
\label{eq:3D_z}
\end{eqnarray}
where $x_{i}$, $y_{i}$, and $z_{i}$ are the dynamical variables of a three-dimensional oscillator. 
The coupling strength is denoted as $J^{(\mu)}$ $(\mu = x,y,z)$. 
The function $F^{(\mu)}(\cdot)$ is a coupling function to other oscillators, which also specifies the nonlinearity of a single oscillator. 
The coefficients $a^{(\mu)}>0$, $b^{(\mu)}$, $c^{(\mu)}$, and $d^{(\mu)}$ are constants. 
The Langevin noise $\eta_{i}^{(\mu)}(t)$ is assumed to be white Gaussian noise, as follows: 
$\langle \eta_{i}^{(\mu)}(t) \rangle =0$ 
and 
$\langle \eta^{(\mu)}_{i}(t)\eta^{(\nu)}_{j}(t^{\prime})\rangle=2D^{(\mu)}\delta_{ij}\delta_{\mu\nu}\delta(t-t^{\prime})$, 
where
$D^{(\mu)}>0$. 
If noise is absent, Eqs. (\ref{eq:3D_x})--(\ref{eq:3D_z}) with $N=1$ recover single-body deterministic dynamics. Here, we take the function $F^{(\mu)}$ in the form of 
\begin{eqnarray}
F^{(x)}(p) &=& F^{(y)}(p) = p, \label{eq:F_xy}\\
F^{(z)}(p) &=& p\displaystyle \exp\left(-\frac{p^{2}}{2}\right), 
\label{eq:F_z}
\end{eqnarray}
which enables us to analytically calculate $\langle F^{(\mu)}\rangle_{\tiny \textrm{G}}$ described later. A limit cycle attractor can be obtained with the appropriately chosen parameter values. In this paper, we consider the situation that (i) $z_{1}$-nullcline takes the Z-curve, which contributes to bistability, (ii) the dynamical variables $x_{1}$ and $y_{1}$ are slow whereas $z_{1}$ is fast, (iii) unstable foci are relevantly observed in the ($x_{1}$, $y_{1}$) plane under singular perturbation \cite{Rssler_Chaotic_1976,Jackson_Perspectives_1990}. The parameter values are close enough to exhibit chaotic behavior, as shown in Fig.~\ref{fig:Fig_3D_dG_DS_BDL}.

Our stochastic model is proposed such that the collective motions of the system can be derived as closed ordinary differential equations through analysis by the nonlinear Fokker--Planck equation (NFPE) \cite{Frank_Nonlinear_2005}. These consist of the order parameters, which include the noise strength as a parameter value of the deterministic nonlinear dynamical system. Thus, investigating the nonlinear aspects of order parameters in terms of the noise strength dependency allows us to study the occurrence of NICS.

Let us suppose the macroscopic behavior of the system. In the thermodynamic limit ($N \rightarrow \infty$), the mean-field coupling terms of Eqs.~(\ref{eq:3D_x})--(\ref{eq:3D_z}) converge owing to the law of large numbers, which attributes to the self-averaging property. Using the empirical probability density 
$P(t, \textrm{\boldmath $u$})$, 
these terms are written as 
\def\bmu{\textrm{\boldmath $u$}}
\def\Fm{\langle F^{(\mu)}\rangle}
\begin{equation}
\displaystyle \Fm \equiv 
\int \textrm{d} \bmu F^{(\mu)}(b^{(\mu)}u^{(x)} + c^{(\mu)}u^{(y)} + d^{(\mu)}u^{(z)})P(t, \bmu), 
\label{eq:MFCoupling1}
\end{equation}
where $\langle \cdot \rangle$ denotes the ensemble average. and $\bmu^{T}=(u^{(x)}, u^{(y)}, u^{(z)})=(x, y, z)$. Based on the self-averaging, Eqs.~(\ref{eq:3D_x})--(\ref{eq:3D_z}) are reduced to a single oscillator written as
\begin{equation}
\displaystyle \dot{u}^{(\mu)} = -a^{(\mu)}u^{(\mu)} + \displaystyle J^{(\mu)} \Fm + \eta'^{(\mu)}(t), 
\label{eq:3D_ReducedLangevinEq}
\end{equation}
where $\eta'^{(\mu)}(t)$ is the white Gaussian noise given by 
$\langle \eta'^{(\mu)}(t) \rangle = 0$ 
and 
$\langle \eta'^{(\mu)}(t)\eta'^{(\nu)}(t^{\prime}) \rangle = 2D^{(\mu)}\delta_{\mu\nu}\delta(t-t^{\prime})$. 
Here, the coupling term is replaced by Eq.~(\ref{eq:MFCoupling1}), which is time dependent. Every oscillator follows Eq.~(\ref{eq:3D_ReducedLangevinEq}) with different initial conditions and $\eta'^{(\mu)}(t)$. The NFPE corresponding to Eq.~(\ref{eq:3D_ReducedLangevinEq}) is obtained as 
\begin{eqnarray}
\displaystyle \frac{\partial}{\partial t}P(t, \textrm{\boldmath $u$})
&=& - \displaystyle \sum_{\mu=x,y,z}\left[ \frac{\partial}{\partial u^{(\mu)}} ( - a^{(\mu)}u^{(\mu)} + J^{(\mu)}\Fm \right. \nonumber \\
&&\mbox{} \left. \displaystyle - D^{(\mu)}\frac{\partial^2}{\partial u^{(\mu) 2}}\right]P. 
\label{eq:NFPE}
\end{eqnarray}
The Gaussian probability density is a special solution for the NFPE.

For sufficiently large times, the H theorem \cite{Shiino_Chaos_2001} ensures that the probability density of Eq.~(\ref{eq:NFPE}) converges to the Gaussian form described as 
\def\bms{\textrm{\boldmath $s$}}
$P_{\tiny \textrm{G}}(t,\bmu)=\exp\left[-\frac{1}{2}\bms^{T}C^{-1}(t)\bms\right]/\sqrt{(2\pi)^3\det C(t)}$, 
where 
$\bms=\bmu-\langle \bmu \rangle$ and 
$C_{\mu\nu}(t)=\langle s^{(\mu)}s^{(\nu)}\rangle$. 
Owing to the characteristics of the Gaussian probability density, Eq.~(\ref{eq:MFCoupling1}) is written in terms of the first and second moments. Hence, one can derive the time evolution of the moments from Eq.~(\ref{eq:NFPE}), which is a closed ordinary differential equation. For the mean, it takes 
\def\xm{\langle x\rangle_{\tiny \textrm{G}}}
\def\ym{\langle y\rangle_{\tiny \textrm{G}}}
\def\zm{\langle z\rangle_{\tiny \textrm{G}}}
\def\dxm{\langle \dot{x}\rangle_{\tiny \textrm{G}}}
\def\dym{\langle \dot{y}\rangle_{\tiny \textrm{G}}}
\def\dzm{\langle \dot{z}\rangle_{\tiny \textrm{G}}}
\def\fmx{\langle F^{(x)}\rangle_{\tiny \textrm{G}}}
\def\fmy{\langle F^{(y)}\rangle_{\tiny \textrm{G}}}
\def\fmz{\langle F^{(z)}\rangle_{\tiny \textrm{G}}}
\begin{eqnarray}
\dxm &=& -a^{(x)}\xm + J^{(x)}\fmx,
\label{eq:3D_MomentEq_mean_x} \\
\dym &=& -a^{(y)}\ym + J^{(y)}\fmy,
\label{eq:3D_MomentEq_mean_y} \\
\dzm &=& -a^{(z)}\zm + J^{(z)}\fmz,
\label{eq:3D_MomentEq_mean_z} 
\end{eqnarray}
and for the variance and covariance, 
\def\vm{\langle v^{(\mu)}\rangle_{\tiny \textrm{G}}}
\def\wmn{\langle w^{(\mu\nu)}\rangle_{\tiny \textrm{G}}}
\def\dvm{\langle \dot{v}^{(\mu)}\rangle_{\tiny \textrm{G}}}
\def\dwmn{\langle \dot{w}^{(\mu\nu)}\rangle_{\tiny \textrm{G}}}
\begin{eqnarray}
\dvm &=&
- 2 a^{(\mu)}\vm + 2D^{(\mu)}, 
\label{eq:3D_MomentEq_var} \\
\dwmn &=&
- ( a^{(\mu)} + a^{(\nu)} )\wmn, 
\label{eq:3D_MomentEq_cov}
\end{eqnarray}
where 
$\langle \cdot \rangle_{\tiny \textrm{G}}$ is expectation over $P_{\tiny \textrm{G}}$, 
\def\umu{\langle u^{(\mu)}\rangle_{\tiny \textrm{G}}}
\def\unu{\langle u^{(\nu)}\rangle_{\tiny \textrm{G}}}
$v^{(\mu)} = (u^{(\mu)}\!-\!\umu)^2$, and 
$w^{(\mu\nu)}=(u^{(\mu)}-\umu)(u^{(\nu)}-\unu)$ ($\mu \neq \nu$). 
Note that 
$\vm \rightarrow D^{(\mu)}/a^{(\mu)}$ and
$\wmn \rightarrow 0$ 
for sufficiently large times, which suggests the existence of three effective dimensions for the order parameters. In addition, the expectation of the coupling function is calculated as 
\begin{eqnarray}
\langle F^{(x)}\rangle_{\tiny \textrm{G}} &=& m^{(x)}, 
\label{eq:3D_SelfAveraging_FdG_x}\\
\langle F^{(y)}\rangle_{\tiny \textrm{G}} &=& m^{(y)}, 
\label{eq:3D_SelfAveraging_FdG_y}\\
\langle F^{(z)}\rangle_{\tiny \textrm{G}} &=& \frac{m^{(z)}}{(\sigma^{2}+1)^{3/2}}\exp\left[-\frac{m^{(z)^{2}}}{2(\sigma^{2}+1)}\right], 
\label{eq:3D_SelfAveraging_FdG_z}
\end{eqnarray}
where 
\def\xg{\langle x\rangle_{\tiny \textrm{G}}}
\def\yg{\langle y\rangle_{\tiny \textrm{G}}}
\def\zg{\langle z\rangle_{\tiny \textrm{G}}}
$m^{(\mu)}=b^{(\mu)}\xg+c^{(\mu)}\yg+d^{(\mu)}\zg$ and 
$\sigma^{2}=b^{(z)^{2}}\langle v^{(x)}\rangle_{\tiny \textrm{G}}+c^{(z)^{2}}\langle v^{(y)}\rangle_{\tiny \textrm{G}}+d^{(z)^{2}}\langle v^{(z)}\rangle_{\tiny \textrm{G}}$. 
The reduced equations (\ref{eq:3D_MomentEq_mean_x})--(\ref{eq:3D_MomentEq_cov}) reflect an aspect of the system as order parameters.

\begin{figure}[t]
  \centering
    \includegraphics[width=8.6cm]{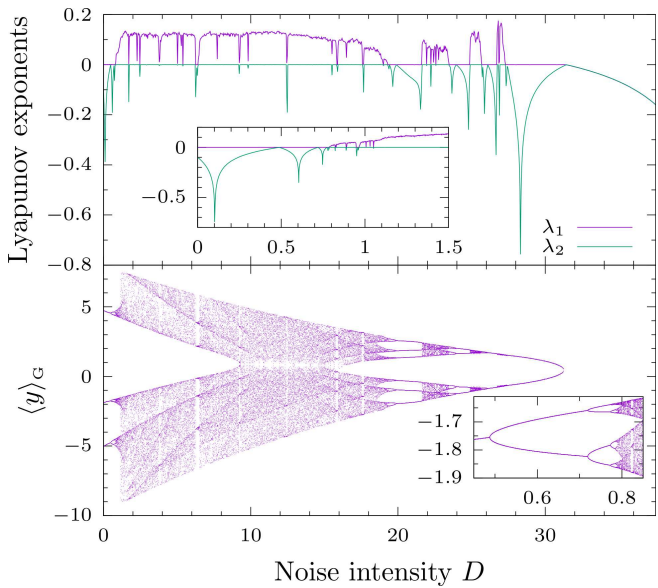}
    \caption{(Color online) Effects of the Langevin noise intensity on the order parameter system. 
    (upper panel) The first and second Lyapunov exponents ($\lambda_1$ and $\lambda_2$). The regions of (i) $\lambda_1>0$, (ii) $\lambda_1=0$ and $\lambda_2<0$, and (iii) $\lambda_1<0$ correspond to chaos, the limit cycles, and the fixed points, respectively. 
    (lower panel) Poincar\'e section. The section is taken as $\xm = 1.38$ and $\zm < 0$. 
    The parameter values are $D^{(x)}=D^{(y)}=D^{(z)}=D$, in addition to those of Fig.~\ref{fig:Fig_3D_dG_DS_BDL}. 
    }
    \label{fig:Fig_3D_dG_LN_LE}
\end{figure}

As described before, the occurrence of NICS is associated with the occurrence of the macroscopic behavior of chaos in the order parameter system. To capture the dependence of attractor types on the Langevin noise intensity, the Lyapunov exponents \cite{Shimada_A_1979} for the system of Eqs.~(\ref{eq:3D_MomentEq_mean_x})--(\ref{eq:3D_SelfAveraging_FdG_z}) were numerically estimated (Fig.~\ref{fig:Fig_3D_dG_LN_LE}). Note that the Lyapunov exponents were measured not for the stochastic differential equations of the oscillators but for the deterministic ordinary differential equations of the order parameters. In the deterministic limit ($D \rightarrow 0$), the system exhibits the limit cycle attractor as well as the single body deterministic case (Fig.~\ref{fig:Fig_3D_dG_DS_BDL}). With increasing noise levels, the attractor become strange, which means the occurrence of NICS in the coupled oscillators. The fixed point attractor appears if the noise level is strong enough, which corresponds to the destruction of the system by noise.

\begin{figure*}[t]
  \centering
    \includegraphics[width=17.2cm]{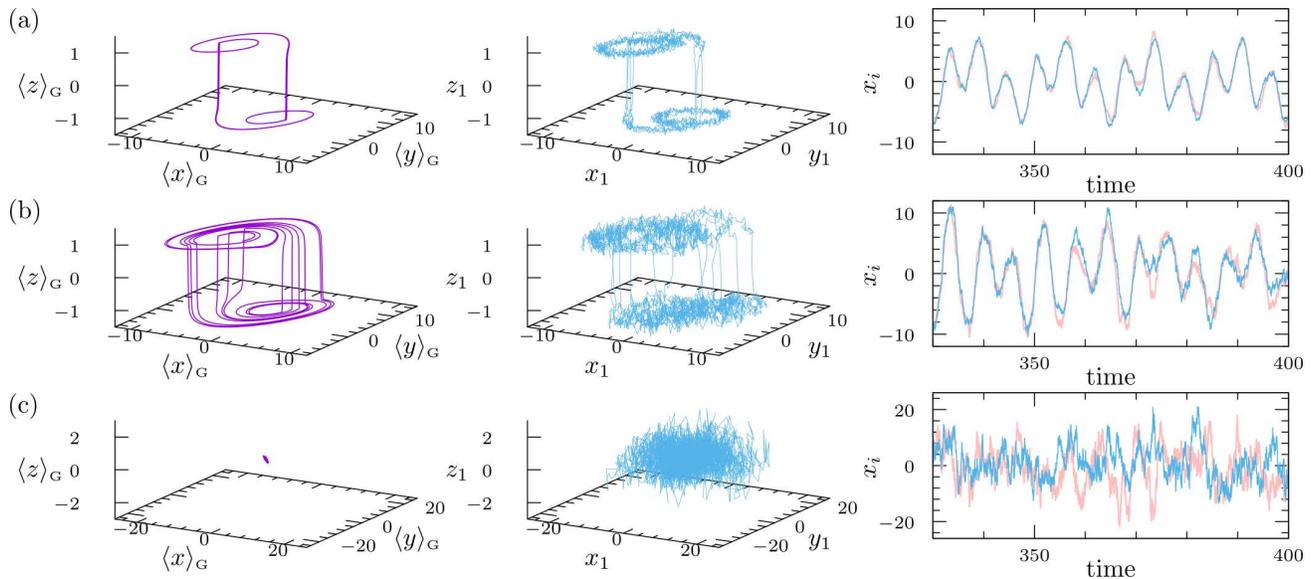}
     \caption{(Color online) An example of dynamical behavior under the presence of noise. 
     The left, center, and right columns show averaged motion in the phase space of the order parameters, a single path in the subspace, and two sample paths in the time course. 
     (a) Deterministic limit ($D=0.3$). The synchronized periodic oscillation is shown. 
     (b) Appropriate noise level ($D=1.5$). The oscillatory behavior is changed by the noise from periodic to chaotic while it is still synchronous. 
     (c) Strong noise level ($D=45.0$). No synchronized oscillation is observed. 
     The system size is $N=10^4$ and the values of the other parameters are the same as those in Fig.~\ref{fig:Fig_3D_dG_LN_LE}. 
     }
     \label{fig:NICSexample}
\end{figure*}

To access the picture of NICS, a numerical simulation for the original stochastic differential equations (\ref{eq:3D_x})--(\ref{eq:F_z}) was accomplished. The Heun method was applied \cite{Kloeden_Numerical_1992}. As shown in Fig.~\ref{fig:NICSexample}, noise could control the dynamical property of the system, which gave rise to the synchronization transition from periodic to chaotic with increasing noise intensity.

So far, a solvable model for noise effects on mean-field coupled limit cycle oscillators has been proposed. One characteristic of this model is that in the thermodynamic limit, the probability density of the system is constrained to the Gaussian form for sufficiently large times due to the independent noise. This enables us to analytically derive the time evolution of the order parameters. Note that the coupling and noise strengths are not limited to weak values in our study. The macroscopic chaotic behavior of the order parameters could imply chaotic synchronization under the noise. Since the noise intensity is one of the control parameters for the time evolution of the order parameter, the noise could induce the chaotic synchronization when the deterministic single oscillator settles near the chaotic parameter values. It should be emphasized again that a solvable model for independent external noise, which describes the occurrence of NICS in coupled limit cycle oscillators, could contribute to resolving the complicated questions that arose after the discovery of NICS in uncoupled oscillators.

Our model would be treated as comprising damped oscillators with the feedback represented by the coupling terms regardless of the microscopic states of the oscillators. It should be noted that we could arbitrarily design the shape of the coupling function, which may yield a rich variety of dynamical behavior. The feedback might correspond to the common environment \cite{Resmi_Synchronized_2010}.

Another aspect of the relationships between chaos and noise effects was studied using solvable models \cite{Shiino_Chaos_2001,Ichiki_Chaos_2007}. These studies focused on the generation of chaotic behavior purely induced by independent noise in globally coupled limit cycle oscillators. It was shown that colored noise and coupling noise can actually induce chaotic behavior in order parameter systems whose effective dimensions include variances. Further investigation into this relationship will be conducted in the future.

Some problems continue to remain unresolved. 
One of these involves phase. If the famous R\"ossler type attractor \cite{Rssler_An_1976} is obtained by changing the parameter values in our model, one can easily define the phase in the systems of both a single oscillator and order parameters. Since synchronization induced by common noise in uncoupled limit cycle oscillators is already theoretically established from the viewpoint of phase \cite{Teramae_Robustness_2004, Nakao_Noise_2007}, a study on NICS in coupled limit cycle oscillators in terms of phase would facilitate systematic understanding of noise-induced synchronization. 
Another open problem involves system symmetry. For instance, the manner of symmetric coupling plays an important role in deterministic chaotic synchronization in coupled oscillators \cite{Pikovsky_Synchronization_2001}. It is expected that symmetric coupling is also the main factor for NICS in our study. These issues will be reported elsewhere.

The authors would like to thank Dr. Shiino for valuable discussions. 
This work was supported by JSPS KAKENHI Grant Numbers JP19K20360, JP17H06469. 



\bibliographystyle{apsrev4-2}

\end{document}